\pdfoutput=1
\documentclass[a4paper,aps,pre,twocolumn,superscriptaddress]{revtex4-1}
\usepackage[utf8]{inputenc}
\usepackage{amsmath}
\usepackage{amsfonts}
\usepackage{amssymb}
\usepackage[english]{babel}
\usepackage{graphicx}
\usepackage{units}
\usepackage{natbib}

\begin{document}

\author{C. Arran}
\affiliation{York Plasma Institute, University of York, York, United Kingdom}
\author{J. M. Cole}
\affiliation{The John Adams Institute for Accelerator Science, Imperial College London, London, United Kingdom}
\author{E. Gerstmayr}
\affiliation{The John Adams Institute for Accelerator Science, Imperial College London, London, United Kingdom}
\author{T. G. Blackburn}
\affiliation{Department of Physics, Chalmers University of Technology, Gothenburg, Sweden}
\author{S. P. D. Mangles}
\affiliation{The John Adams Institute for Accelerator Science, Imperial College London, London, United Kingdom}
\author{C. P. Ridgers}
\affiliation{York Plasma Institute, University of York, York, United Kingdom}

\title{Optimal Parameters for Radiation Reaction Experiments}

\begin{abstract}
As new laser facilities are developed with intensities on the scale of $\unit[10^{22}-10^{24}]{W cm^{-2}}$, it becomes ever more important to understand the effect of strong field quantum electrodynamics processes, such as quantum radiation reaction, which will play a dominant role in laser-plasma interactions at these intensities. Recent all-optical experiments, where GeV electrons from a laser wakefield accelerator encountered a counter-propagating laser pulse with $a_0 > 10$, have produced evidence of radiation reaction, but have not conclusively identified quantum effects nor their most suitable theoretical description. Here we show the number of collisions and the conditions required to accomplish this, based on a simulation campaign of radiation reaction experiments under realistic conditions. We conclude that while the critical energy of the photon spectrum distinguishes classical and quantum-corrected models, a better means of distinguishing the stochastic and deterministic quantum models is the change in the electron energy spread. This is robust against shot-to-shot fluctuations and the necessary laser intensity and electron beam energies are already available. For example, we show that so long as the electron energy spread is below $25\%$, collisions at $a_0 = 10$ with electron energies of $\unit[500]{MeV}$ could differentiate between different quantum models in under $30$ shots, even with shot to shot variations at the $50\%$ level.
\end{abstract}


\maketitle

\section{Introduction}


Experiments using new ultra-high-intensity multi-petawatt laser facilities such as Apollon 10 PW \cite{zou2015} and ELI \cite{weber2017,gales2018} will require a thorough experimental understanding of non-classical behavour in laser-plasma interactions. In experiments reaching laser intensities of $\unit[10^{22}-10^{24}]{Wcm^{-2}}$ the effect of strong field QED processes starts to strongly modify the laser-plasma interaction \cite{bell2008,dipiazza2012,mourou2006}, and better understanding the fundamental physical processes at work will be crucial. One of these processes, the radiation produced by charged particles when moving in an electro-magnetic field, and the subsequent recoil experienced by the particles, is particularly relevant to studies of inverse Compton scattering \cite{powers2013,tsai2016} and laser absorption in solid target interactions \cite{duff2018,tamburini2010,zhang2015,wang2017}, both of which are key targets of the ELI-NP facility.

\begin{figure}
\begin{center}
\includegraphics[scale=0.5,trim=0cm 12cm 5cm 0cm]{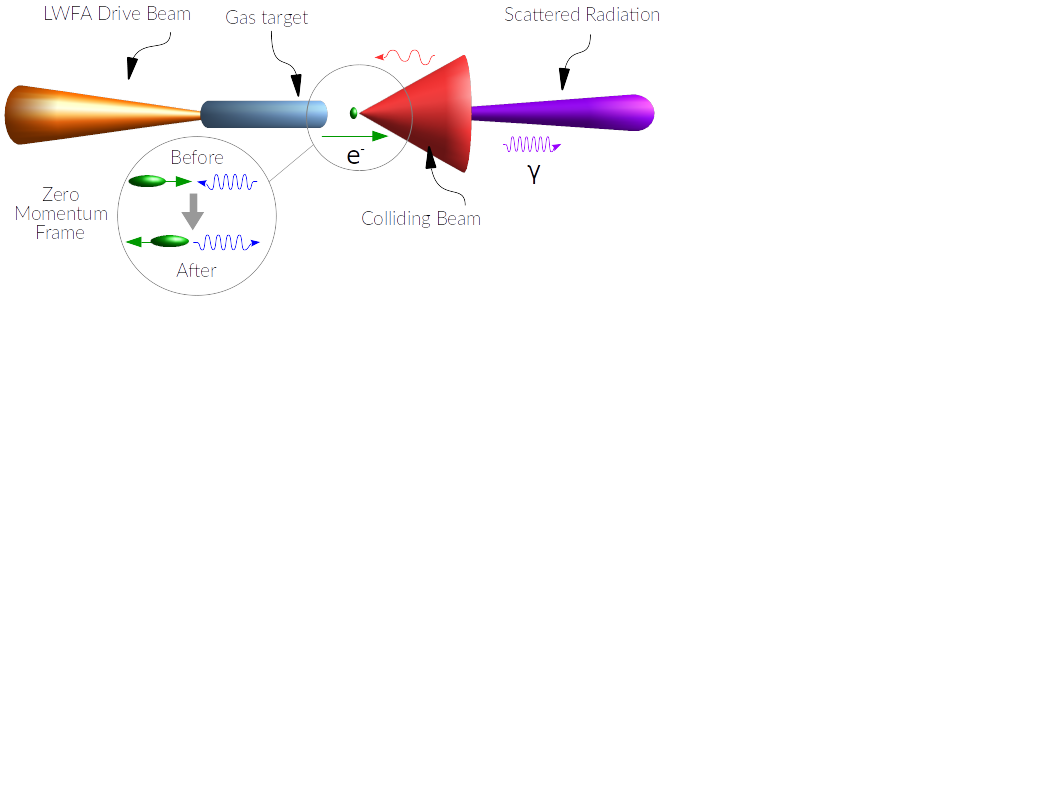}
\caption{Schematic of an all-optical Radiation Reaction experiment. An intense ultra-short laser pulse, the LWFA drive beam, is incident upon a gas target, producing a high energy electron bunch. A second intense laser pulse, the colliding beam, is brought to a tight focus just outside the gas target, interacting with the electron bunch and producing a beam of high energy light.}
\label{fig: RR Schematic}
\end{center}
\end{figure}

Two recent experiments \cite{cole2018,poder2018a} have aimed to study the effect of radiation reaction in isolation, using existing petawatt laser facilities with peak intensities of ${I \sim \unit[10^{21}]{Wcm^{-2}}}$. In these all-optical set ups, highly energetic electrons ($\gamma = 1000 - 2000$) produced by a laser wakefield accelerator \cite{tajima1979,mangles2004,faure2004,geddes2004} collide with a counter-propagating high-intensity laser pulse as shown in the schematic in Fig.~\ref{fig: RR Schematic}. In the rest frame of the electron both the frequency and the intensity of the radiation are dramatically increased, bringing the electric field experienced by the electron to the scale of ${E'_L \approx \gamma E_L \sim \unit[10^{17}]{V/m}}$, comparable to the Schwinger limit ${E_s = \unit[1.32 \times 10^{18}]{V/m}}$ \cite{schwinger1951}, as described by the dimensionless and Lorentz invariant parameter ${\chi_e = E'_L / E_s}$. At this point the predictions from quantum and classical models of radiation reaction strongly diverge; whereas using the classical synchrotron spectrum requires the production of photons with energies $\varepsilon_\gamma > \varepsilon_e$, the quantum model limits the energy of photons so as to conserve energy, significantly reducing the synchrotron power at high field strengths, as described in refs.~\cite{erber1966,sokolov2010,ridgers2014,blackburn2014}. Both of the recent experiments demonstrate significantly better agreement between their measurements and quantum, non-perturbative, models than with classical models such as described by Landau and Lifshitz \cite{landau1971}.

However, the limited number of events measured in the experiments has left significant uncertainty \cite{macchi2018}, with Poder \emph{et al.} \cite{poder2018a} concluding a slightly better agreement with a semi-classical model, while the measurements made by Cole \emph{et al.} \cite{cole2018} were not able to distinguish between the semi-classical and stochastic models. In the semi-classical description, both the rate of radiation emission and the subsequent change in electron energy are adjusted to match the quantum model, but the emission remains a continuous process, with the recoil a frictional force that leads to cooler electrons with a narrower energy distribution. In the quantum picture, on the other hand, emission is a quantized, stochastic event; some electrons travel much further through the laser pulse before emitting a photon while others emit many, leading to substantial broadening of the electron energy distribution \cite{neitz2013,ridgers2017,niel2018}. In modelling stochastic emission events, we assume that photon emission is sufficiently fast that the laser field is constant throughout the process, in the so-called constant-cross-field approximation. This is accurate when the coherence time of emission is much less than the laser period, which generally gives a condition $\tau_\mathrm{COH} \sim m c / e E_L \ll 1/\omega $, for a laser frequency $\omega$ \cite{ritus1985}, or in terms of the normalized vector potential of the laser pulse, $a_0 = e E_L / m_e \omega c \gg 1$. Even if this condition is met, however, the constant-crossed-field approximation -- and both the quantum and semi-classical models -- breaks down when the energy of the emitted photon energy is very low \cite{harvey2015,dipiazza2018}, although these photons do not contribute significantly to the recoil \cite{blackburn2018}.

Experiments currently underway seek to resolve the seeming disparity between the two experiments to date and to determine which, if either, model is most appropriate for high intensity laser experiments; this paper attempts to both find the best way of conducting these radiation reaction experiments, and demonstrate the regimes where the choice of model is important. By simulating the experiments under different conditions we place constraints on the parameters required, such as laser intensity and electron energy spread, as well as the accuracy to which these parameters must be controlled. Given different experimental parameters, we estimate the number of measurements required to be confident which model is more appropriate: the stochastic quantum model, or the continuous and deterministic semi-classical model. In doing so, we take account of the shot-to-shot variation of both the energy of electrons from a laser wakefield accelerator and the intensity of the colliding laser pulse.

\section{Simulated Experiments}


In a radiation reaction experiment of the type shown in Fig.~\ref{fig: RR Schematic} it is important to achieve good overlap in both time and space between the highest intensity region of the laser pulse and the brightest part of the electron beam. However, if the pulse collides with the electron beam close to the LWFA gas target, the electron bunch will be under one micron in diameter. If the collision point is slightly away from the focal plane of the laser it is possible to ensure that the beam profile is much larger than the electron bunch, maximising the overlap in space and the chances of a successful collision. Under these conditions we can reduce the problem to a single dimension. Similarly, synchrotron radiation is emitted within a forward-pointing cone with an angle of $1/\gamma$ around the direction of motion of the electron; for an electron bunch with an angular divergence on the scale of $\theta \sim \unit[1]{mrad}$ the total cone angle will be $\approx \theta + a_0 / \gamma \sim \unit[10]{mrad}$ in the plane of polarization of the laser. Under these conditions we will ignore the angular distribution of radiation for energetic electrons. On the short timescale after the creation of the electron beam we can also neglect direct electron-electron interactions such as space-charge, while the electron bunch duration is sufficiently short that we can neglect the interactions of synchrotron photons after they have been emitted. Likewise, as the number of electrons in the bunch is small, we neglect effects of the electron bunch on the laser beam, such as energy loss and refraction. Finally, as we will be considering situations achievable with existing laser facilities, with both $E_L/E_s \ll 1$ and $\chi_e \lesssim 1$, we can neglect pair production. We therefore consider the interaction between each electron and the laser pulse independently, allowing us to reduce a complicated simulation to a sum of many single particle interactions, where each electron has a single initial and final energy but may produce many photons over the course of the interaction with the laser pulse.

First, look-up tables were assembled of the final electron and photon spectra resulting when initially monoenergetic electron beams encountered a laser pulse. One dimensional Particle-In-Cell simulations with EPOCH \cite{arber2015}, employing an extended QED module \cite{ridgers2014} (see appendix~A for details), were conducted for laser peak intensities of $1 \leq a_0 \leq 25$ and electron energies of $100 \leq \varepsilon_i \leq \unit[2]{GeV}$. These used each in turn of a fully classical, Landau-Lifshitz, radiation reaction model; a semi-classical model with corrected emission rates and powers but with continuous, deterministic emission; and a quantum model of radiation reaction with stochastic emission. In each simulation, the laser pulse had a Gaussian profile with a duration of $\unit[40]{fs}$ FWHM, chosen to reflect parameters of the recent radiation reaction experiments.

\begin{figure*}
\begin{center}
\includegraphics[scale=0.35,trim=1cm 0 0 0]{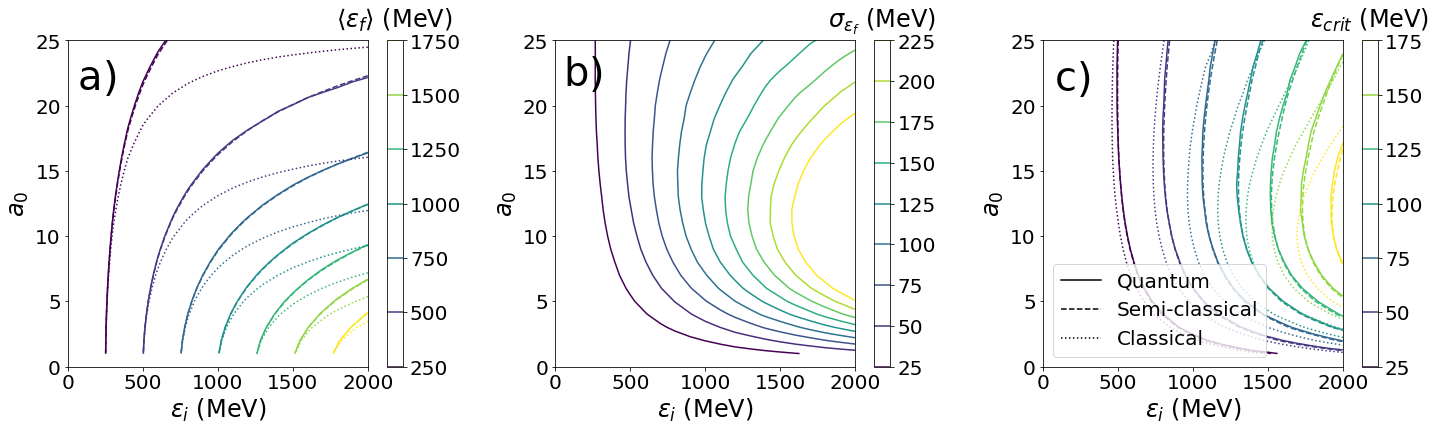}
\caption{Contours of a) the mean final electron energy $\langle \varepsilon_f \rangle$ b) the final electron energy spread $\sigma_{\varepsilon_f}$ and c) the critical energy of emitted photons $\varepsilon_\mathrm{crit}$, for each of the quantum (solid lines), semi-classical (dashed) and classical (dotted) models. Results are from mono-energetic electron beams in EPOCH simulations. Each line shows the initial electron energy $\varepsilon_i$ and the laser $a_0$ required to obtain the given final state. For the classical and semi-classical models $\sigma_{\varepsilon_f} < \unit[1]{MeV}$ and hence the contours are not visible.}
\label{fig: Look-up Table Contours}
\end{center}
\end{figure*}

For each different reaction model, these look-up tables gave final electron and photon energy distributions $N_{e,f}(\varepsilon_f | \varepsilon_i,a_0)$ and $N_\gamma(\varepsilon_\gamma | \varepsilon_i,a_0)$. The electron energy distributions were fitted to a Gaussian to give functions $\langle \varepsilon_f \rangle(\varepsilon_i,a_0)$ and $\sigma_{\varepsilon_f}(\varepsilon_i,a_0)$, whereas the photon energy distributions were fitted to an expression of the form:
\begin{equation}
N_\gamma \propto \varepsilon_\gamma^{-\frac{2}{3}} \exp \left( - \frac{\varepsilon_\gamma}{\varepsilon_\mathrm{crit}} \right), \label{eqn: Ecrit}
\end{equation}
giving the critical energy $\varepsilon_\mathrm{crit}(\varepsilon_i,a_0)$. For examples of the fitted energy distributions, see appendix~\ref{sec: Monte-Carlo}. The resulting parameters are plotted in Fig.~\ref{fig: Look-up Table Contours} using contour plots to show the differences between the three different models.

Firstly, Fig.~\ref{fig: Look-up Table Contours} demonstrates that the simulations of radiation reaction are working as expected. For a given initial electron energy, increasing the laser intensity reduces the final energy of the electrons, while increasing the initial electron energy leads to the emission of higher energy photons. Both of these correspond to a larger radiation reaction force, with an electron beam losing more power, emitted as photons. However, although at low values of $a_0$ the photon critical energy increases with laser intensity, for higher $a_0$ it saturates and for the highest values of $\varepsilon_i$ actually begins to decrease. This is because electrons lose so much energy during the radiation reaction process that the peak field experienced by the electrons in their rest frame, $\gamma E_L$, is actually reduced. In this situation the radiated spectrum comprises a greater number of lower energy photons. Similarly, in the quantum model, electrons experience the greatest stochastic broadening at moderate laser intensities, around $a_0 \approx 10$, while above this the final energy spread is smaller. At moderate laser intensities electrons emit fewer photons on average, with greater variation between electrons due to shot noise.

The look-up tables also allow us to distinguish between the different models for radiation reaction: as laser intensity increases, the classical model predicts much lower final electron energies than the quantum or semi-classical models. Applying the quantum correction -- limiting the photon energy to $\varepsilon_\gamma < \varepsilon_e$ -- results in significantly higher final electron energies and slightly lower photon energies. The greatest difference in final electron energy occurs at the highest laser intensites and electron energies, whereas the greatest difference in critical energy is centered around $a_0 \approx 10$.

In both the mean final electron energy and photon energy, it is very difficult to see any difference between the predictions from the quantum and semi-classical models. These models contain the same correction to the power radiated, and on average the electrons encounter the same radiation reaction. However, without the effect of stochastic broadening, an initially mono-energetic electron beam remains mono-energetic in the semi-classical model. In contrast, in the quantum model the electron energy spectrum becomes substantially broader.

Once the look-up tables $\langle \varepsilon_f \rangle (\varepsilon_i,a_0)$, $\sigma_{\varepsilon_f}(\varepsilon_i,a_0)$ and $\varepsilon_\mathrm{crit}(\varepsilon_i,a_0)$ were assembled, Monte-Carlo simulated experiments were conducted for more realistic (though still idealised) initial electron energy distributions, which were not mono-energetic, and where each of the laser intensity, the mean electron energy, and the electron energy spread varied shot-to-shot. This allowed us to estimate the underlying 3-dimensional probability distribution function $f(\langle \varepsilon_f \rangle, \sigma_{\varepsilon_f},\varepsilon_\mathrm{crit})$ for making measurements of the final mean electron energy $\langle E_f \rangle$, and energy spread $\sigma_{E_f}$, and the final photon critical energy $\varepsilon_\mathrm{crit}$. For parameter scans, the 2-dimensional probability distribution functions $f_1(\langle \varepsilon_f \rangle,\varepsilon_\mathrm{crit})$ and $f_2(\langle \varepsilon_f \rangle, \sigma_{\varepsilon_f})$ were calculated instead. The details are described in appendix~B.

\section{Distinguishing Models}

\begin{figure*}
\begin{center}
\includegraphics{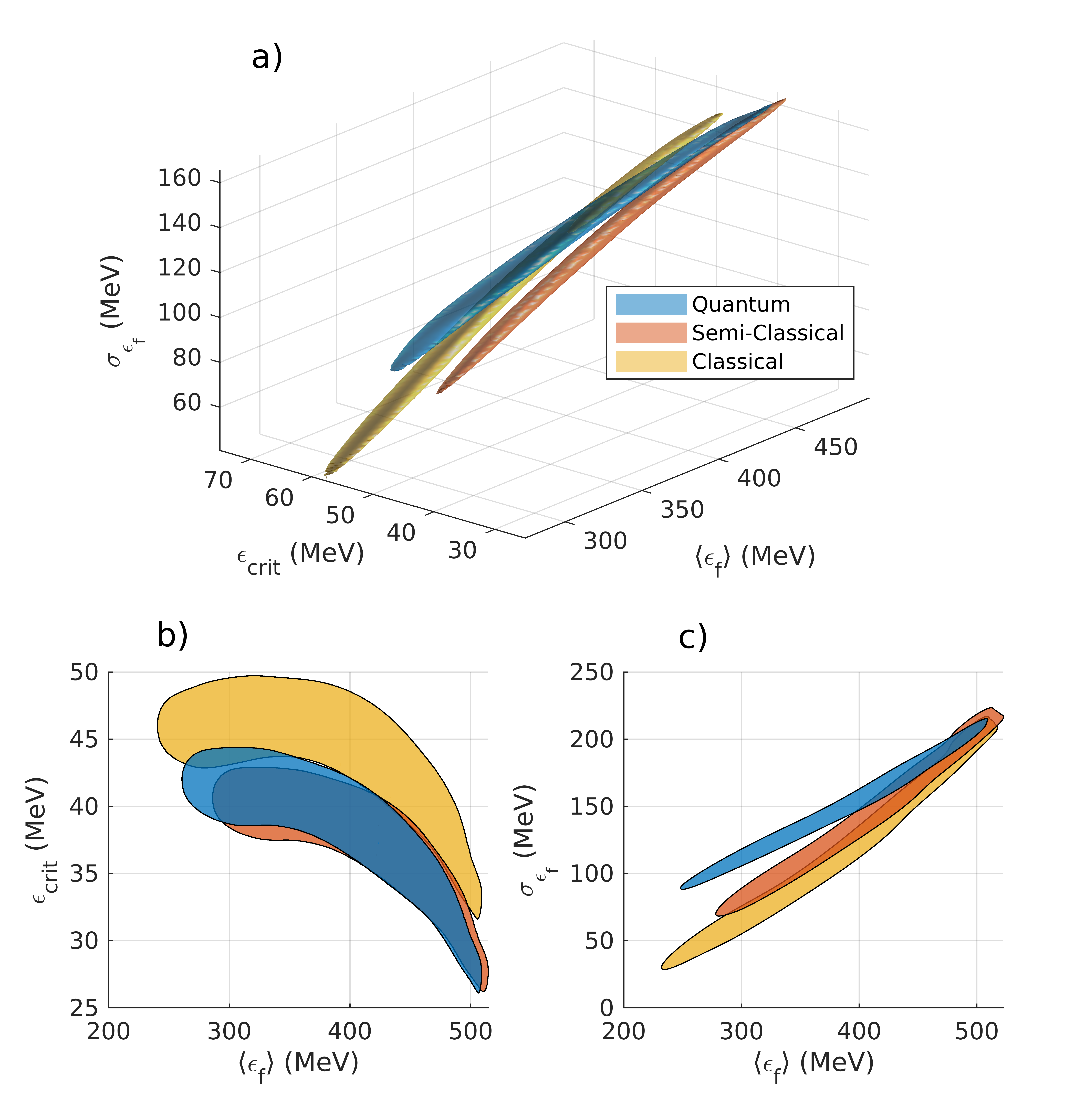}
\caption{Results from simulated radiation reaction experiments for an initial electron beam with a peak energy of $\unit[(550 \pm 20)]{MeV}$ and an energy spread of $\unit[250]{MeV}$, shown through a) the 3-dimensional joint probability distribution function $f(\langle \varepsilon_f \rangle,\varepsilon_\mathrm{crit},\sigma_{\varepsilon_f})$, and the 2D distribution functions b) $f_1(\langle \varepsilon_f \rangle,\varepsilon_\mathrm{crit})$ and c) $f_2(\langle \varepsilon_f \rangle,\sigma_{\varepsilon_f})$. The $1\sigma$ contours are shown, within which $68\%$ of simulated experiments measured these results, assuming each of the classical (yellow), semi-classical (red), and quantum (blue) models. In a) the intensity of the colliding laser pulse is $a_0 = 11 \pm 3$, whereas for b) and c) the intensity was assumed uniformly distributed between $a_0 = 4 - 20$, for comparison with Fig.~9 of ref.\cite{cole2018}.}
\label{fig: PRX Parameters}
\end{center}
\end{figure*}

First, experimental parameters were chosen to match those in ref.~\cite{cole2018}, with the laser intensity estimated as $a_0 = 11 \pm 3$ and the electron energy estimated as $\langle \varepsilon_i \rangle = \unit[(550 \pm 20)]{MeV}$, with an energy spread of $\sigma_{\varepsilon_i} = \unit[250]{MeV}$. The 3-dimensional distribution functions, shown in Fig.~\ref{fig: PRX Parameters}, demonstrate the capability of the simulated experiments. Points in the top right of the image correspond to shots with low $a_0$, with high final energies, high energy spread, and lower photon energies. In this regime, the three models predict very similar results. As $a_0$ increases, the electron beam loses more energy and becomes cooler with a lower energy spread, in the process producing higher energy photons. At the largest values of $a_0$, the different rates of radiation reaction and of radiative cooling lead to the three models predicting different results, with the classical model leading to the lowest final energy spread and the highest photon energy, while the quantum model predicts a significantly higher energy spread than either of the two other models. The shot to shot variation of $\langle \varepsilon_i \rangle$ tends to blur out this trend, broadening the distribution functions and making it more difficult to distinguish between different models.

In order to show this more clearly, and to compare the results with ref.~\cite{cole2018}, the 2-dimensional distribution functions $f_1$ and $f_2$ were calculated for laser intensities pulled from a uniform distribution between $a_0 = 4$ and $a_0 = 20$. $f_1(\langle \varepsilon_f \rangle,\varepsilon_\mathrm{crit})$, shown in Fig.~\ref{fig: PRX Parameters}b), agrees well with the previous work, with the classical model predicting substantially higher critical energies as expected. The predictions from the quantum and semi-classical models strongly overlap, however, and hence using critical energy from the photon spectra is a poor way of determining between stochastic and semi-classical models of radiation reaction.

Fig.~\ref{fig: PRX Parameters}c), however, shows another possible measurement using the same experimental parameters, comparing the mean final electron energy with the final energy spread, as described by $f_2(\langle \varepsilon_f \rangle,\sigma_{\varepsilon_f})$. Using these measurements the semi-classical and classical models predict fairly similar results, but the quantum model predicts a substantially higher energy spread than both of the other models. This is because, for an interaction where the electron beam is much smaller than the focal spot, these deterministic models predict that electron beams can only ever become cooled by emitting synchrotron radiation in the electric field of the laser, as more energetic electrons emit radiation more strongly. In the stochastic model, however, the number and energy of photons emitted by each electron is probabilistic and varies strongly. In certain cases, the final energy spread of the electron may increase over time (see for instance refs.~\cite{vranic2016,ridgers2017,niel2018}). The lower the energy spread, the more likely this becomes. Ref.~\cite{ridgers2017} eqn. (3.8) predicts that for Gaussian energy spectra the cross-over point, below which the stochastic broadening dominates, can be approximated in the case $\sigma_{\varepsilon_i} \ll \langle \varepsilon_i \rangle$ by:
\begin{align}
\left( \frac{\sigma_{\varepsilon_i}}{\langle \varepsilon_i \rangle}\right)^2 &\approx \frac{55\sqrt{3}}{8 \cdot 24} \langle \chi_e \rangle \\
\frac{\sigma_{\varepsilon_i}}{\langle \varepsilon_i \rangle} &\approx 0.70 \sqrt{\langle \chi_e \rangle} \label{eqn: Broadening cross-over}
\end{align}

We can quantify the ease of distinguishing between models by measuring the overlap between the joint distribution functions. The probability of making a measurement of several parameters, denoted by the vector $\mathbf{x}$, given that a model $A$ is true, is denoted $P(\mathbf{x}|A)$. The chances of incorrectly inferring model $B$ from those measurements can be related by Bayes' theorem to the model probability as $P(B|\mathbf{x}) = P(\mathbf{x}|B)P(B)/P(\mathbf{x})$. If the prior assumption is that the two models are equally likely, but not the only two possible models, and that all measurements of $\mathbf{x}$ in the region of interest are equally likely, with no bias in the measuring equipment, we can show that the probability of incorrectly inferring model $B$ from a single measurement of $\mathbf{x}$, if model $A$ is true (or vice versa) is proportional to the overlap $\Omega$ between the models, as:
\begin{equation}
P(B|A) = P(A|B) = \Omega \equiv \frac{\int P(\mathbf{x}|A) \cdot P(\mathbf{x}|B) \mathrm{d} V_\mathbf{x}}{\sqrt{\int P(\mathbf{x}|A)^2 \mathrm{d} V_\mathbf{x}\int P(\mathbf{x}|B)^2 \mathrm{d} V_\mathbf{x}}},
\end{equation}
where the integrations are performed over the domain of possible measurements within which $P(\mathbf{x})$ is constant. This probability is normalized such that if $P(\mathbf{x}|A)=P(\mathbf{x}|B)$ then $\Omega=1$. Depending on the choice of measurements, $P(\mathbf{x}|A)$ and $P(\mathbf{x}|B)$ are described by the joint distribution functions $f_1(\langle \varepsilon_f \rangle, \varepsilon_\mathrm{crit})$ and $f_2(\langle \varepsilon_f \rangle, \sigma_{\varepsilon_f})$ for each model.

If $N$ independent and identically distributed measurements are taken, one for each successful laser-beam collision, the probability of incorrectly inferring model $B$, given that $A$ is in fact true, becomes $P(B|A) = \Omega^N$. Conversely, if we require better than a certain degree of accuracy to be sure we will not incorrectly infer model $B$, such that $P(B|A) < p$, we can show that we require $N > \log p / \log \Omega$. If the overlap between the joint distribution functions is very high, $\Omega \rightarrow 1$, $|\log \Omega | \rightarrow 0$ and it becomes increasingly difficult to confirm which model is correct, requiring an ever larger number of shots.

This is shown in Table~\ref{tab: PRX Overlaps} for the parameters described in ref.~\cite{cole2018} and joint distribution functions shown in Fig.~\ref{fig: PRX Parameters}. The classical model predicts significantly different results to the quantum and semi-classical models and hence the overlap and number of shots required are both small. The work in ref.~\cite{cole2018} was therefore able to show that the quantum model agreed better with the data than the classical model, despite only definitely measuring four successful collisions. In general, measurements of $\varepsilon_\mathrm{crit}$ and $\langle \varepsilon_f \rangle$ are successful at determining between classical and quantum/semi-classical models of radiation reaction. With those measurements, however, it would be more difficult to distinguish between the quantum and semi-classical models, with at least 70 shots required to obtain the same level of certainty.

An alternative approach is to measure $\sigma_{\varepsilon_f}$, therefore significantly reducing both the overlap between the models and the number of shots required. Although an accurate measurement of $\sigma_{\varepsilon_f}$ is difficult, requiring a clean electron energy spectrum, the difference in the predictions from quantum and semi-classical models is significant. For certain experimental parameters, using this measurement could reduce by more than an order of magnitude the number of shots required to confidently determine which model is more appropriate. 

\begin{table*}
\begin{tabular}{|c|c c |c c|}
\hline 
 & \multicolumn{2}{c|}{$\Omega$} & \multicolumn{2}{c|}{$N_{min}$ for $p=0.3\%$} \\ 
\hline 
Models & $f_1(\langle \varepsilon_f \rangle, \varepsilon_\mathrm{crit})$ & $f_2(\langle \varepsilon_f \rangle, \sigma_{\varepsilon_f})$ & \hspace{0.75cm} $f_1$ \hspace{0.75cm} & \hspace{0.75cm} $f_2$ \hspace{0.75cm} \\ 
\hline 
Quantum   / Classical      & 0.235 & 0.134 & 4 & 3 \\ 
Quantum   / Semi-Classical & 0.920 & 0.241 & 70 & 4 \\ 
Classical / Semi-Classical & 0.180 & 0.446 & 4 & 7 \\ 
\hline 
\end{tabular}
\caption{Overlap $\Omega$ of joint distribution functions $f_1$ and $f_2$ and corresponding minimum number of shots required to obtain $3\sigma$ confidence in determining between models for radiation reaction, using different sets of measurements.}
\label{tab: PRX Overlaps}
\end{table*}

\section{Optimal Parameters}

\begin{figure*}
\begin{center}
\includegraphics{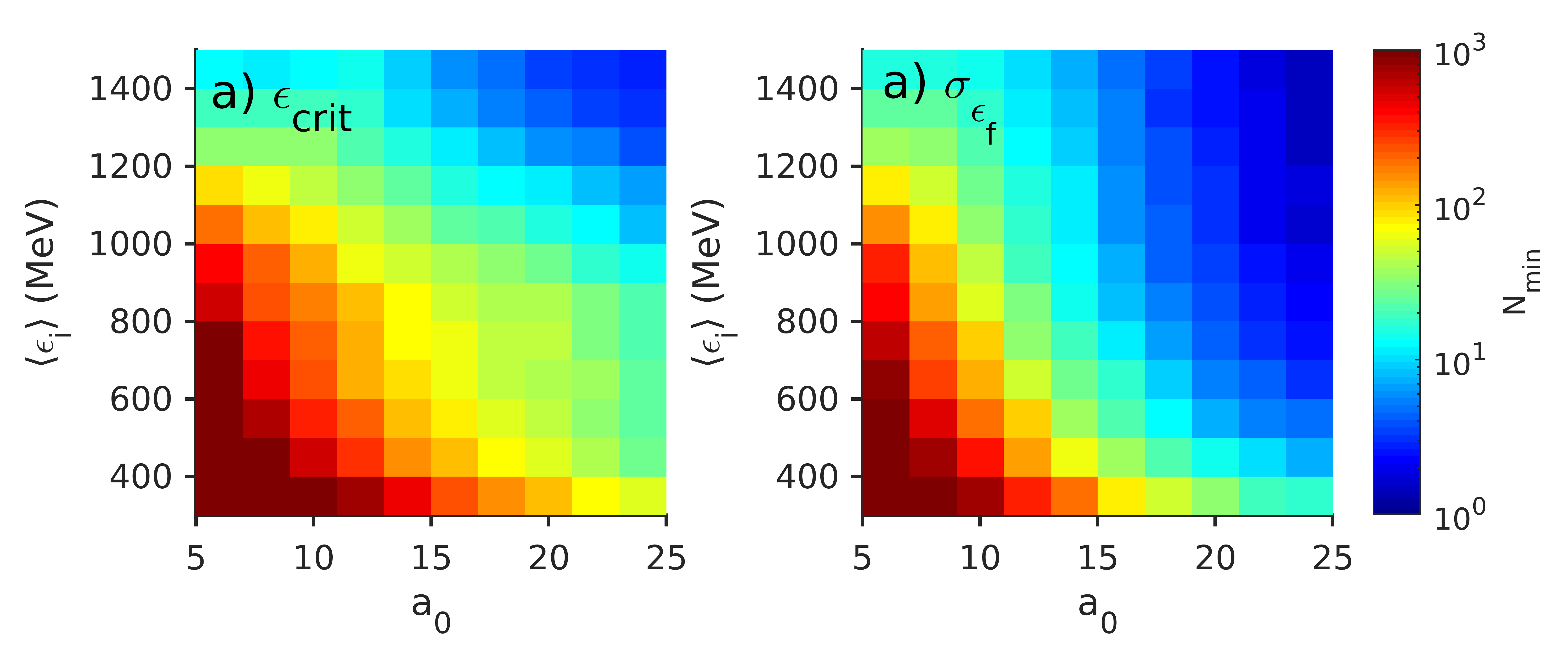}
\caption{The estimated number of shots required to distinguish between the quantum and semi-classical models at the $p=0.3\%$ significance level using measurement of a) $\langle \varepsilon_f \rangle$ and $\varepsilon_\mathrm{crit}$ and b) $\langle \varepsilon_f \rangle$ and $\sigma_{\varepsilon_f}$, plotted against the laser intensity $a_0$ and the electron energy $\varepsilon_i$. The variation on $a_0$ was taken to be $\pm 3$, and the shot-to-shot variation on $\langle \varepsilon_i \rangle$ was $\pm 10\%$. The initial energy spread was $\sigma_{\varepsilon_i} = (0.5 \pm 0.125) \langle \varepsilon_i \rangle$. The colour scale is the same for both plots.}
\label{fig: PRX Parameter Scan}
\end{center}
\end{figure*}

The overlap between the joint distribution functions assuming quantum and semi-classical models of radiation reaction was tabulated over a wide range of different initial electron energies and laser intensities in order to determine the number of shots required to distinguish at the $p=0.3\%$ level between the two models given certain experimental parameters. Similar values as before for the uncertainties in $a_0$ and $\langle \varepsilon_i \rangle$ were used, at $\pm 3$ and $\pm 10\%$ respectively, with a very large energy spread of $\sigma_{\varepsilon_i} / \langle \varepsilon_i \rangle = 50\%$ as before. In order to describe realistic experiments, shot-to-shot variation on $\sigma_{\varepsilon_i}$ has also been introduced, at $\pm 25\%$ of $\sigma_{\varepsilon_i}$, such that the laser intensity, mean electron energy, and electron energy spread all vary shot to shot. The results are shown in Fig.~\ref{fig: PRX Parameter Scan}.


Fig.~\ref{fig: PRX Parameter Scan}a) shows the number of shots required when using measurements of $\langle \varepsilon_f \rangle$ and $\varepsilon_\mathrm{crit}$; this demonstrates that when operating at realistic experimental parameters, many shots must be taken to distinguish between quantum and semi-classical models of radiation reaction. For low $a_0$ and $\langle \varepsilon_i \rangle$, hundreds of shots are required to conclude that one of the models is correct and not the other. In contrast to the situation for mono-energetic electron beams, increasing the laser intensity above $a_0 = 10$ increases the difference between the predictions from the two models. If it is possible to increase the electron energy to $\langle \varepsilon_i \rangle \geq \unit[1]{GeV}$ and the laser intensity to $a_0 \geq 15$, the number of shots required is reduced to below 25. Under these conditions, a practical radiation reaction experiment could determine with significant ($>3\sigma$) confidence which model is more appropriate in this regime.

If we cannot determine between quantum and semi-classical models of radiation reaction using measurements of $\langle \varepsilon_f \rangle$ and $\varepsilon_{crit}$, it is possible to use a measurement of the final electron energy spread $\sigma_{\varepsilon_f}$. As shown in Fig.~\ref{fig: PRX Parameter Scan}b), this does not reduce the number of shots required at low laser intensities, but is substantially more successful at lower electron energies, with fewer than 25 shots required if $\langle \varepsilon_i \rangle \geq \unit[500]{MeV}$ and $a_0 \geq 15$. For sufficiently high electron energies and laser intensities (for instance $\langle \varepsilon_i \rangle \geq \unit[1]{GeV}$ and $a_0 \geq 20$, the predictions from the quantum and semi-classical models for radiation reaction are substantially different; under these conditions, it is vital to understand which model of radiation reaction is more accurate, and only a single shot may be sufficient to discriminate between quantum and semi-classical models.

\begin{figure*}
\begin{center}
\includegraphics{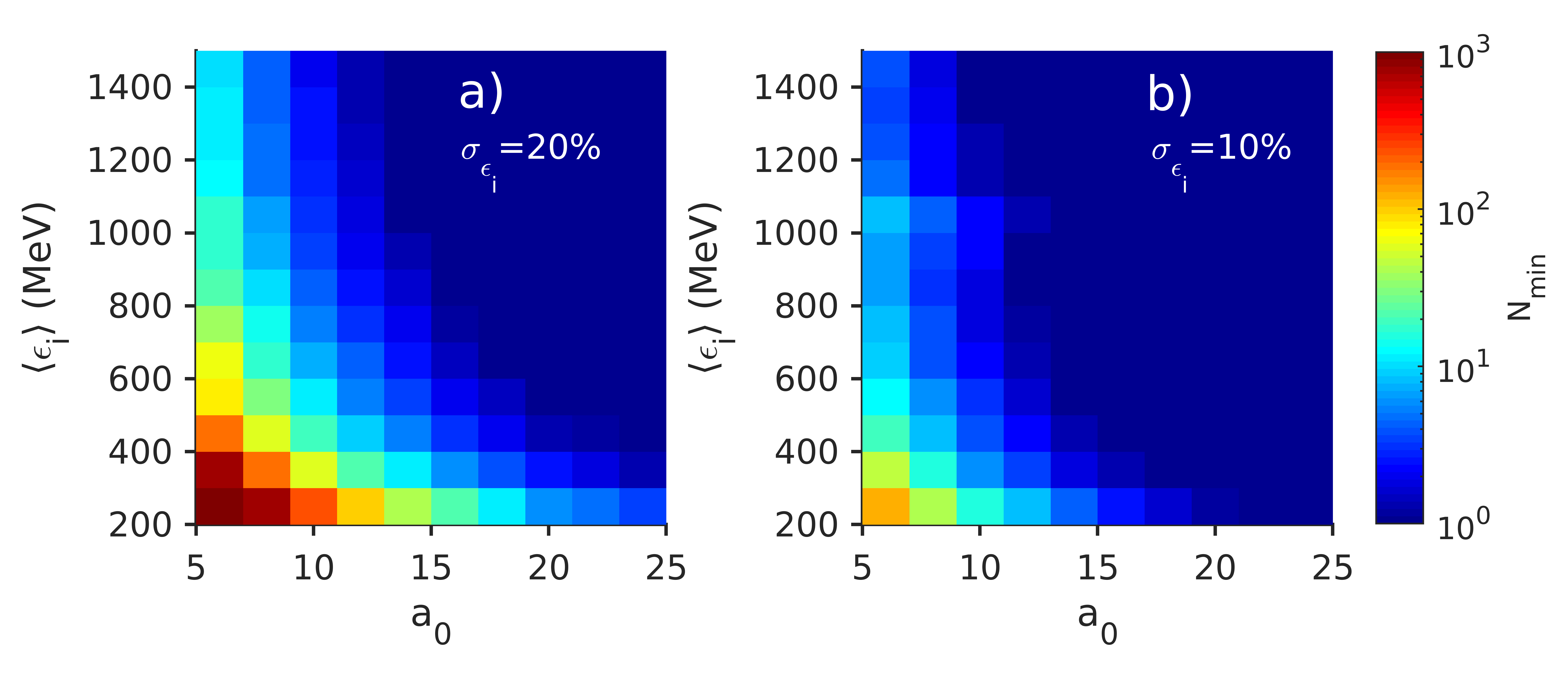}
\caption{Using an initial energy spread of a) $\sigma_{\varepsilon_i} / \langle \varepsilon_i \rangle=20\% \pm 5\%$ and b) $\sigma_{\varepsilon_i} / \langle \varepsilon_i \rangle=10\% \pm 2.5\%$, the estimated number of shots required to distinguish between quantum and semi-classical models of radiation reaction at a confidence level of $p=0.3\%$. All values are using measurements of $\langle \varepsilon_f \rangle$ and $\sigma_{\varepsilon_f}$ and are plotted against the laser intensity $a_0$ and the electron energy $\varepsilon_i$. The shot-to-shot variation on $a_0$ was taken to be $\pm 3$, and the variation of $\varepsilon_i$ was $\pm 10\%$. The colour scale is the same for both plots.}
\label{fig: Low Energy Spread Scan}
\end{center}
\end{figure*}

It is clear that the energy spread of the electron spectrum is the key distinguishing feature of a stochastic model as opposed to a deterministic model, and we can study the effect of decreasing the initial energy spread in the electron spectrum. This will reduce the cooling experienced by the electron beam, increasing the relative contribution of stochasticity. Simulated experiments were run as before, but with the relative energy spread reduced to $20\%$ and $10\%$. Again, the relative variation of the energy spread was $\pm 25\%$, giving energy spreads of $\sigma_{\varepsilon_i} = (0.2 \pm 0.05) \langle \varepsilon_i \rangle$ and $(0.1 \pm 0.025) \langle \varepsilon_i \rangle$, respectively. As shown in Fig.~\ref{fig: Low Energy Spread Scan}, reducing the initial energy spread significantly changes the final energy spread and reduces the number of shots required to determine which model is more suitable. If no more than one hundred shots are possible and the relative energy spread is $20\%$, distinguishing between the models requires just $\langle \varepsilon_i \rangle \geq \unit[500]{MeV}$ or $a_0 \geq 12$. Under most conditions simulated, fewer than 10 shots would be required. For an energy spread of $10\%$, however, the models can easily be distinguished even for the lowest laser intensities and electron energies. Only at $\langle \varepsilon_i \rangle = \unit[200]{MeV}$ and $a_0 = 5$ do the models predict very similar outcomes; under these conditions the accuracy of the constant-cross-field approximation is doubtful and it is likely that both models will break down. In most of the simulated experiments, however, only a single shot would be sufficient to determine which model is more correct. Reducing the energy spread of the initial electron beam is therefore one of the best ways of ensuring an experiment will be able to distinguish between deterministic and stochastic models of radiation reaction.

\begin{figure*}
\begin{center}
\includegraphics{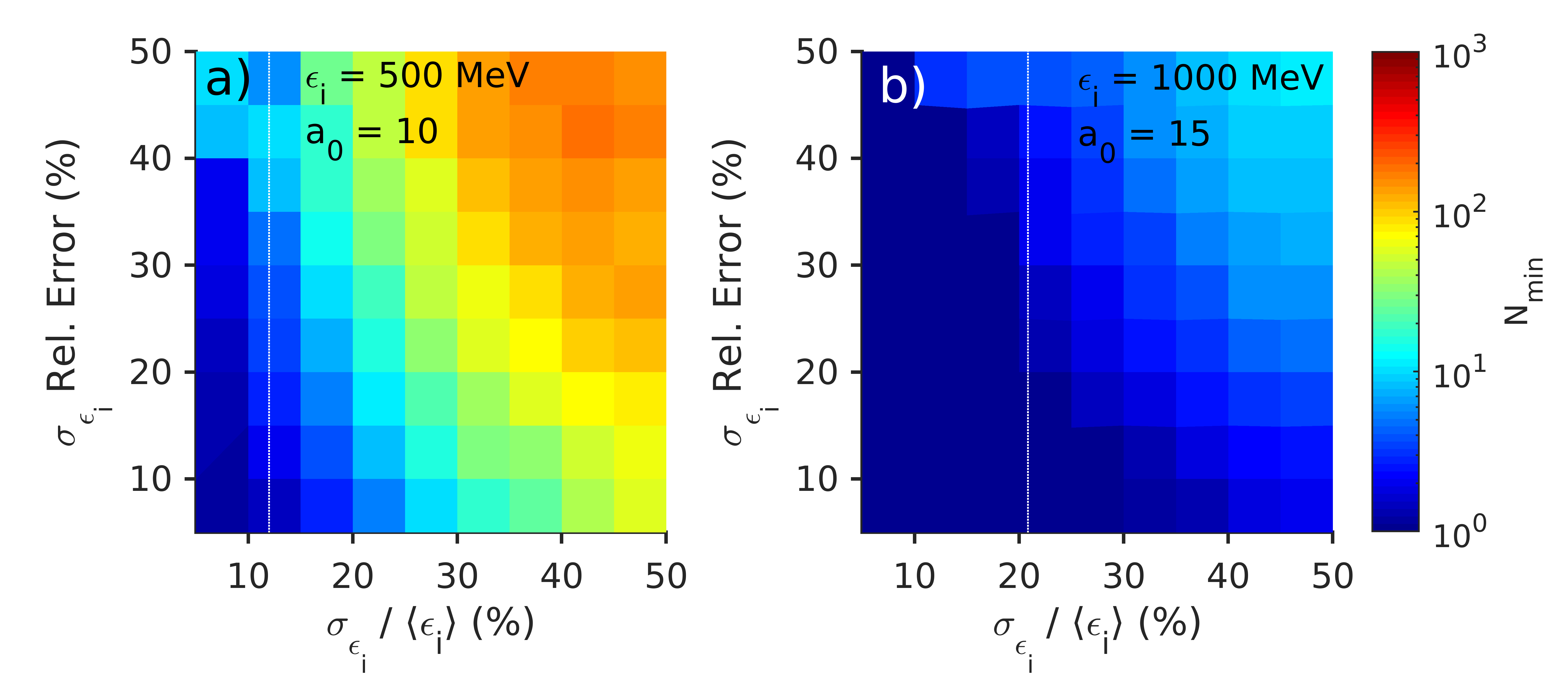}
\caption{Using initial conditions of a) $a_0 = 10 \pm 3$ and $\langle \varepsilon_i \rangle = \unit[(500 \pm 50)]{MeV}$, and b) $a_0 = 15 \pm 3$ and $\langle \varepsilon_i \rangle = \unit[(1.0 \pm 0.1)]{GeV}$, the number of shots required for Monte-Carlo simulated experiments to determine with $3\sigma$ confidence between the quantum and semi-classical models. All values are using measurements of $\langle \varepsilon_f \rangle$ and $\sigma_{\varepsilon_f}$ and are plotted against the electron energy spread, relative to the mean electron energy, and the shot-to-shot variation of the electron energy spread, relative to the energy spread. The colour scale is the same for both plots. The analytic prediction to measure stochastic broadening in the ideal case is also shown (white dotted line).}
\label{fig: Energy Spread Error Scan}
\end{center}
\end{figure*}

We can study the maximum allowable energy spread at a certain laser intensity and initial electron energy, if stochastic effects are to be measured. First, we ran simulated experiments for $a_0 = 10 \pm 3$ and ${\langle \varepsilon_i \rangle = \unit[(500 \pm 50)]{MeV}}$, which corresponds to a quantum parameter ${\langle \chi_e \rangle = \langle \gamma_e E_L / E_s \rangle \approx 0.03}$. These parameters are achievable in many existing PW scale laser facilities, and are on the same scale as achieved previously in ref.~\cite{cole2018}. The results, plotted in Fig.~\ref{fig: Energy Spread Error Scan}a), show the number of shots required for a range of energy spreads $\sigma_{\varepsilon_i}$ and errors on the energy spread. The number of shots required increases with both the energy spread and the variation in the energy spread, and is sufficient to make experiments impractical when either of these are high. For an experiment to be practical, requiring only 10s of shots to successfully distinguish between models, the energy spread must generally be kept below around $\sigma_{\varepsilon_i} / \langle \varepsilon_i \rangle \lesssim 25\%$. If the relative error on the energy spread can be greatly reduced, however, experiments with these parameters can be successful while $\sigma_{\varepsilon_i} / \langle \varepsilon_i \rangle \leq 50\%$. For $\langle \chi_e \rangle \approx 0.03$ in eqn.~\eqref{eqn: Broadening cross-over}, the energy spread required for the electron spectrum to broaden is $\sigma_{\varepsilon_i} / \langle \varepsilon_i \rangle \lesssim 12 \%$, below which simulated experiments measure a clear difference between quantum and semi-classical models. However, the simulated experiments show that a significant difference arises between the two models well before stochastic broadening dominates, so long as the variation on the energy spread is limited to a few tens of percent or lower.

We also ran simulated experiments for $a_0 = 15 \pm 3$ and $\langle \varepsilon_i \rangle = \unit[(1.0 \pm 0.1)]{GeV}$, or $\langle \chi_e \rangle \approx 0.1$, parameters which are around the limit of what is achievable with some of the existing petawatt laser facilities (e.g. \cite{poder2018a}). Fig.~\ref{fig: Energy Spread Error Scan}b) shows the results, demonstrating that the number of shots required is greatly reduced at these parameters. Under these conditions, it is relatively straightforward to distinguish between the quantum and semi-classical models, with the predicted energy spread significantly different even when the energy spread is on the level of $\sigma_{\varepsilon_i} / \langle \varepsilon_i \rangle \approx 50\%$ and varies widely shot-to-shot. Again, we can calculate where stochastic broadening dominates analytically using eqn.~\eqref{eqn: Broadening cross-over}, giving a condition on the energy spread of $\sigma_{\varepsilon_i} / \langle \varepsilon_i \rangle \lesssim 20 \%$. With this increased laser intensity and electron energy, the simulations show that the predictions of the two models diverge even well above this threshold, with the stochastic model predicting a large reduction in the cooling rate and a significantly different final energy spread.

\begin{figure*}
\begin{center}
\includegraphics{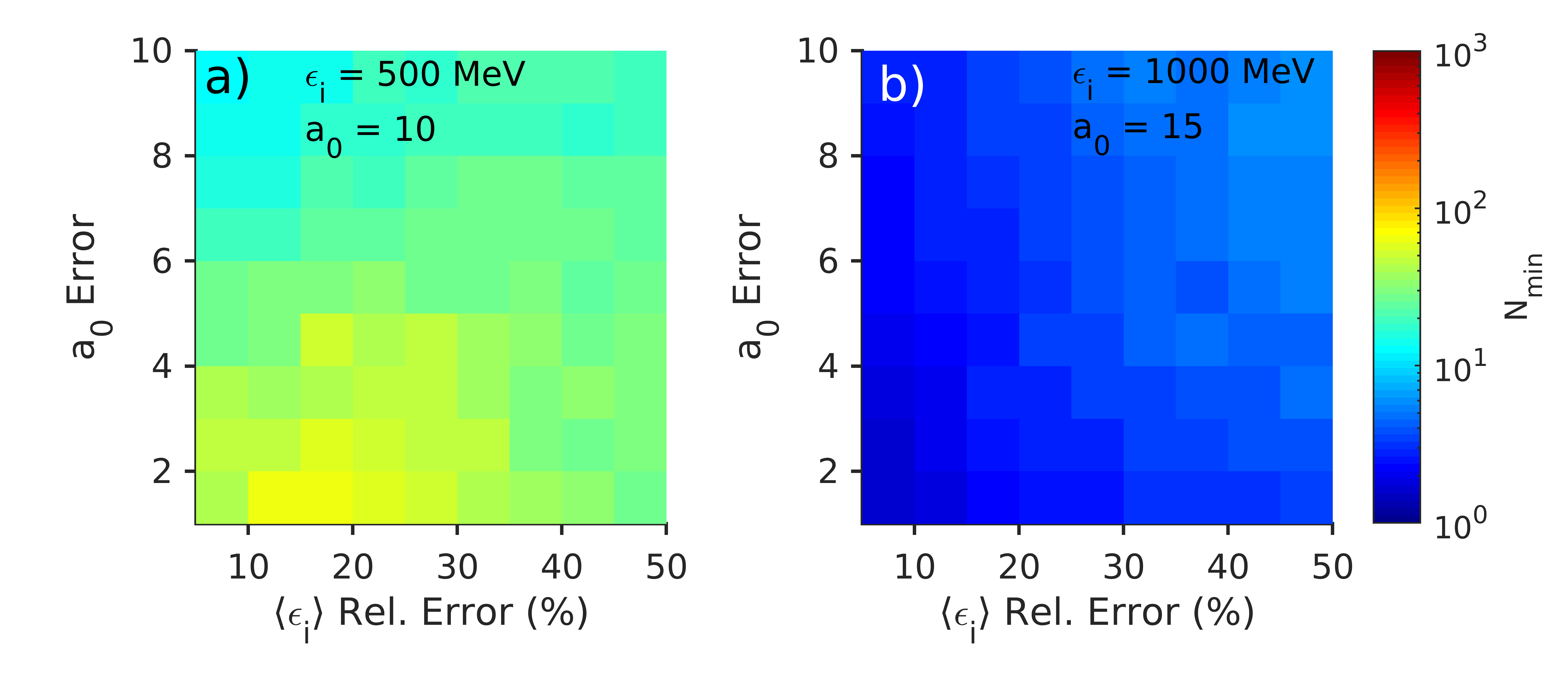}
\caption{For a) $a_0 = 10$ and $\langle \varepsilon_i \rangle = \unit[500]{MeV}$, and b) $a_0 = 15$ and $\langle \varepsilon_i \rangle = \unit[1]{GeV}$, the estimated number of shots required to distinguish at the $p=0.3\%$ significance level between the quantum and semi-classical models. All values are using measurements of $\langle \varepsilon_f \rangle$ and $\sigma_{\varepsilon_f}$ and are plotted against the relative shot-to-shot variation on the initial electron energy and the variation on the laser intensity. The colour scale is the same for both plots.}
\label{fig: Mean Energy Error Scan}
\end{center}
\end{figure*}

Finally, we explored the effect of shot-to-shot variation of the laser intensity and mean initial electron energy under the same two sets of conditions: $a_0 = 10$ and $\langle \varepsilon_i \rangle = \unit[500]{MeV}$; and $a_0 = 15$ and $\langle \varepsilon_i \rangle = \unit[1]{GeV}$. The energy spread was set as $\sigma_{\varepsilon_i} = 0.25 \langle \varepsilon_i \rangle$, with the shot-to-shot variation on the energy spread $0.25 \sigma_{\varepsilon_i}$. Fig.~\ref{fig: Mean Energy Error Scan} shows the number of shots required at a range of errors on both $a_0$ and $\langle \varepsilon_i \rangle$. Interestingly, the effect is small, with no drastic change in the overlap between predictions from the two different models. At a higher laser intensity and electron energy the number of shots required increases slowly with an increasing shot-to-shot variation, as expected. At the lower intensity and electron energy, however, large variation actually results in more significant radiation reaction effects in the high energy and high intensity tails, causing a slight reduction in the number of shots required. Experiments will remain able to distinguish between stochastic and deterministic models of radiation reaction even if the shot-to-shot variation in the laser intensity and electron energy are high, so long as the shot-to-shot variation in electron energy spread is limited.

In the course of these simulated experiments we have considered a wide range of experimental errors, but the laser pulse profile and the electron energy distribution have remained idealised. In practice, electron bunches from LWFA often contain a significant lower energy or thermal component, particularly when operating in the so-called `bubble' regime \cite{pukhov2004}, where electrons are continuously injected into the wakefield. This work has neglected that background, which would have to be carefully removed from the energy spectrum before measuring the final energy spread of the beam. Spatial variation in both the electron beam and the laser pulse can also result in significant changes to the final electron energy spread; if one region of the electron beam experiences a much higher laser intensity and a greater radiation reaction force, the final energy spread of the beam can be significantly higher than expected. Practical experiments must work to limit the spatial variation, which can be achieved by moving the focal plane of the laser pulse further away from the point of collision with the electron beam, at the cost of reducing the effective laser intensity. Future simulated experiments, on the other hand, could include variation of the laser intensity within a single shot, as well as shot-to-shot \cite{baird2018}.



\section{Conclusions}


We have used a series of particle-in-cell simulations and Monte-Carlo simulated experiments to make predictions for radiation reaction from each of the quantum, classical, and semi-classical models using realistic parameters. In doing so, we have shown that while measurements correlating the critical energy of the resulting photon spectra with the mean final electron energy give a way of clearly distinguishing the classical model from the quantum and semi-classical models, this is a poor way of determining which of the quantum and semi-classical models to use. For laser intensities $a_0 < 15$ and electron energies $\varepsilon_i < \unit[1]{GeV}$, these two models predict almost the same final average energy of electrons and photons, for both mono-energetic electron beams and more realistic broad electron distributions.

Instead, we have shown that measuring the energy spread of the electron spectrum after the interaction gives a clearer distinction between the stochastic and deterministic models. Although the energy spread will only increase over the course of the interaction when the initial energy spread is very low (around $\sigma_{\varepsilon_i} / \langle \varepsilon_i \rangle \leq 12\%$ for $a_0 = 10$ and $\varepsilon_i = \unit[500]{MeV}$), the effect of stochasticity substantially reduces the rate of cooling, leading to different predictions from the quantum and semi-classical models even at much higher energy spreads. 

We have used the simulated experiments to determine how many shots would be required when operating at certain conditions, and used this to build up a picture of the optimal experimental parameters. Crucially electron energy spread should be reduced to below $25\%$, if possible, to maximise the chances of conclusively determining which model is more accurate, while shot-to-shot variation on the energy spread should be minimised. At this energy spread, it should be possible to distinguish the quantum and semi-classical models in a few shots even with relatively unambitious experimental parameters, such as $\langle \varepsilon_i \rangle = \unit[500]{MeV}$ and $a_0 = 10$. Under these conditions, the measurement is robust to significant shot-to-shot variation in electron energy and laser intensity, even on the scale of $50\%$. 

Alternatively, if it is difficult to reduce the electron energy spread, increasing the electron energy or laser intensity will separate the predictions from the two models. Even with an energy spread of $\sigma_{\varepsilon_i} = 50\%$, it is possible to distinguish between the models fairly clearly if $a_0 > 10$ and $\langle \varepsilon_i \rangle > \unit[1]{GeV}$, or if $a_0 > 15$ and $\langle \varepsilon_i \rangle > \unit[500]{MeV}$. By an electron energy of $\langle \varepsilon_i \rangle = \unit[1]{GeV}$ and a laser intensity of $a_0 = 15$, the models can be fairly easily distinguished regardless of a large energy spread or shot-to-shot variation. These requirements are certainly achievable using current laser systems, and upcoming experiments should be able to clearly determine which model of radiation reaction is most suitable for describing interactions of energetic electrons with high intensity lasers.

\begin{acknowledgments}

C.A. and C.P.R. are grateful for EPSRC grant no. EP/M018156/1 which made this work possible. This project has received funding from the European Research Council (ERC) under the European Union's Horizon 2020 research and innovation programme (grant agreement no. 682399), from EPSRC grant no. EP/M018555/1, and from the Knut and Alice Wallenberg Foundation. Simulations used the EPOCH PIC code developed under UK EPSRC grants EP/G054950/1, EP/G056803/1, EP/G055165/1 and EP/ M022463/1, while computing resources were provided by STFC Scientific Computing Department's SCARF cluster.

\end{acknowledgments}

\bibliography{library}

\newpage

\onecolumngrid
\appendix

\section{EPOCH Revisions} \label{sec: EPOCH Revisions}

The quantum electro-dynamics model in EPOCH described in detail in ref.~\cite{ridgers2014} is a Monte-Carlo stochastic model which takes into account both the changes to the synchrotron spectrum and emission rate as the effective electric field rises, and also the random nature of the photon emissions. A charged macro-particle is initialised with a random optical depth $\tau_0$. Its optical depth is then reduced at every timestep by:
\begin{equation}
\tau_n = \tau_{n-1} - \frac{\mathrm{d} N_\gamma}{\mathrm{d} t} \cdot \delta t,
\end{equation}
where $\tau_n$ is the optical depth at the $n$-th timestep and $\delta t$ is the duration of the timestep. When the optical depth falls below zero, the macro-particle emits a single macro-photon, with a particle weight equal to that of the original macro-particle and an energy chosen at random from the relevant synchrotron spectrum, as:
\begin{align}
\varepsilon_\gamma &= 2 m_e c^2 \gamma_e \frac{\chi_\gamma}{\chi_e} \\
w_\gamma &= w_e,
\end{align}
where $\chi_\gamma$ is chosen from the synchrotron spectrum $P_{\chi_\gamma}(\chi_\gamma | \chi_e)$, where $\chi_e = \gamma_e E_L / E_s$, where $\gamma_e = 1 + \varepsilon_e / m_e c^2$ is the usual relativistic factor, and $E_s \approx \unit[1.32 \times 10^{18}]{V/m}$ is the Schwinger limit.

We wish, however, to explore two alternative deterministic models: the fully classical model, which possesses neither the random nature of emission nor the changes to the synchrotron spectrum and emission rate; and the so-called semi-classical model, which contains the changes to the synchrotron spectrum and rate, but not the random emission. In these two models, each macro-electron now emits a macro-photon at every timestep, ignoring the optical depth. The energy is chosen at random from the relevant synchrotron spectrum as before (with the classical model using the limit of the synchrotron spectrum as $\chi_e \rightarrow 0$), but the particle weight of this macro-photon is now proportional to the instantaneous emission rate and the timestep duration as:
\begin{align}
w_\gamma &= w_e \cdot \frac{\mathrm{d} N_\gamma}{\mathrm{d} t} \cdot \delta t
\end{align}

In this way, the semi-classical model will predict the same photon energy spectrum and rate of emission as the quantum model, but the semi-classical model is deterministic, with no element of randomness. In this model, charged particles continually emit photons.

For vanishing small $E_L \rightarrow 0$ both the particle weight and energy of the macro-photon should vanish to zero and emission under these conditions will contribute negligibly to the final photon spectrum. There is, however, an additional complexity due to the implementation of $P_{\chi_\gamma}(\chi_\gamma | \chi_e)$ in EPOCH. This is tabulated, and has a lower limit as $\chi_e$ vanishes to zero with $E_L\rightarrow 0$, at a value $\chi_{e,\mathrm{min}}$. This implies that as $\chi_e$ vanishes to zero, $\chi_\gamma$ does not, and so $\varepsilon_\gamma \rightarrow \infty$, which is clearly unphysical.

As $\chi_e \rightarrow 0$, $P_{\chi_\gamma}(\chi_\gamma | \chi_e)$ should instead tend towards the classical limit, where the synchrotron spectrum is a function of a single variable only: $P_{\chi_\gamma}(\chi_\gamma | \chi_e) \rightarrow P(\chi_\gamma / \chi_e^2)$. For $\chi_e < \chi_{e,\mathrm{min}}$, we instead calculate the photon energy using:
\begin{equation}
\chi_\gamma = \chi_\gamma' \left( \frac{\chi_e }{\chi_{e,\mathrm{min}}} \right)^2,
\end{equation}
where $\chi_\gamma'$ is chosen at random from $P_{\chi_\gamma}(\chi_\gamma' | \chi_{e,\mathrm{min}})$. In this way, even though $\chi_\gamma'$ cannot vanish to zero, both $\chi_\gamma$ and $\varepsilon_\gamma \propto \chi_\gamma / \chi_e$ will safely vanish to zero as $\chi_e \rightarrow 0$.

This step is not generally necessary in the stochastic case, as the probability of emitting a macro-photon vanishes to zero as $E_L \rightarrow 0$, so photons with un-physically high energies are never created. When emitting a macro-photon at every timestep, however, this step is important to avoid a large population of extremely high energy macro-photons, even though the particle weights of these macro-photons safely vanish to zero.

\section{Monte-Carlo Simulated Experiments}~\label{sec: Monte-Carlo}

\begin{figure*}[h!]
\begin{center}
\includegraphics{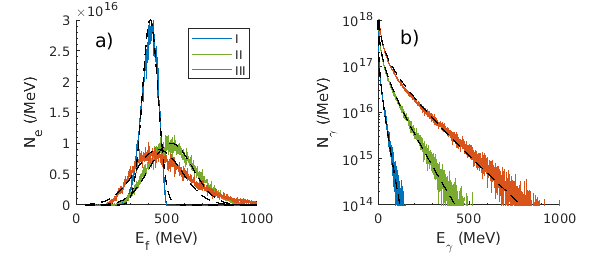}
\caption{Example spectra for a) electrons and b) photons resulting from simulations of monoenergetic electron beams interacting with laser pulses. Electron spectra were fitted to Gaussian distributions (dashed black lines), whereas photon spectra were fitted to eqn.~\eqref{eqn: Ecrit}. Conditions were I) a low laser intensity and electron energy ($a_0=10$, $E_i = \unit[500]{MeV}$, II) a moderate intensity and energy ($a_0=15$, $E_i = \unit[1000]{MeV}$), and III) a high intensity and energy ($a_0=20$, $E_i = \unit[1500]{MeV}$).}
\label{fig: Fitted Spectra}
\end{center}
\end{figure*}

For each simulated shot, values for the laser intensity, parameterised by $a_0^{(n)}$, the mean initial electron energy $\langle \varepsilon_i \rangle^{(n)}$, and the initial energy spread $\sigma_{\varepsilon_i}^{(n)}$ were randomly chosen from Gaussian distributions with a chosen mean and standard deviation. $N_\mathrm{electron} = 10,000$ electrons were then simulated, each with an initial energy $\varepsilon_i^{(s)}$ sampled from another Gaussian distribution, with mean $\langle \varepsilon_i \rangle^{(n)}$ and standard deviation $\sigma_{\varepsilon_i}^{(n)}$. For each simulated electron, the final energy distribution was characterised by a Gaussian with mean $\langle \varepsilon_f \rangle(\varepsilon_i^{(s)},a_0^{(n)})$ and standard deviation $\sigma_{\varepsilon_f}(\varepsilon_i^{(s)},a_0^{(n)})$ drawn from the look-up table, and the final electron energy was then estimated by drawing a random sample $\varepsilon_f^{(s)}$ from this distribution. Example electron energy spectra from the monoenergetic simulations are shown in Fig.~\ref{fig: Fitted Spectra}a) alongside the Gaussian fits. When the electron energy and laser intensity are very high, the final electron spectra are strongly skewed and the Gaussian distribution becomes a worse approximation, but at the laser intensities and electron energies considered in this paper the divergence is small.

A histogram was assembled for each shot, using the $N_\mathrm{electron}$ different values of $\varepsilon_f^{(s)}$, and this was fitted to a Gaussian to estimate $\langle \varepsilon_f \rangle^{(n)}$ and $\sigma_{\varepsilon_f}^{(n)}$. At the same time, for each electron the photon distribution $N_\gamma(\varepsilon_\gamma | \varepsilon_i^{(s)},a_0^{(n)})$ was calculated from the look-up table and added to a total distribution $N_{\gamma}^{(n)}(\varepsilon_\gamma | N_{e,i}^{(n)}, a_0^{(n)})$. This was then fitted to eqn.~\eqref{eqn: Ecrit} to give an estimate $\varepsilon_\mathrm{crit}^{(n)}$. Examples of photon spectra from the mono-energetic simulations are shown in Fig.~\ref{fig: Fitted Spectra}b) alongside fits to eqn.~\eqref{eqn: Ecrit}. Summing the photon spectra means that, as in real experiments, the much higher number of photons emitted by the highest energy electrons tend to dominate the spectrum and using $\varepsilon_\mathrm{crit}^{(n)}$ remains a reasonable way of parameterising the measured spectrum.

In total, $N_\mathrm{shots}=10,000$ shots were simulated, and the different estimates of $\langle \varepsilon_f \rangle^{(n)}$, $\sigma_{\varepsilon_f}^{(n)}$ and $\varepsilon_\mathrm{crit}^{(n)}$ were combined using a Gaussian kernel-density estimate to form the joint distribution function $f(\varepsilon_\mathrm{crit}, \langle \varepsilon_f \rangle, \sigma_{\varepsilon_f})$. The process was repeated for each of the classical, semi-classical and quantum models of radiation reaction, using the appropriate look-up tables, giving three different joint distribution functions $f^{(Q)}$, $f^{(S)}$, and $f^{(C)}$. For parameter scans, however, it is computationally expensive to calculate the 3-dimensional distribution function, and so the 2-dimensional distribution functions $f_1(\langle \varepsilon_f \rangle, \varepsilon_\mathrm{crit})$ and $f_2(\langle \varepsilon_f \rangle, \sigma_{\varepsilon_f})$ were used. These are effectively integrations of $f(\varepsilon_\mathrm{crit}, \langle \varepsilon_f \rangle, \sigma_{\varepsilon_f})$, integrated over $\sigma_{\varepsilon_f}$ or $\varepsilon_\mathrm{crit}$ respectively, and require just $N_\mathrm{shots}=1,000$ shots to accurately sample the underlying distribution.  Convergence testing, varying $N_\mathrm{shots}$, allowed us to estimate the error on the probability under these conditions as approximately $5\%$.

\end{document}